# Low Carrier Density Metal Realized in Candidate Line-Node Dirac Semimetals CaAgP and CaAgAs

Yoshihiko Okamoto[1,3,*], Takumi Inohara[1], Ai Yamakage[1,3], Youichi Yamakawa[2,3], and Koshi Takenaka[1]

[1]*Department of Applied Physics, Nagoya University, Nagoya 464-8603, Japan*
[2]*Department of Physics, Nagoya University, Nagoya 464-8602, Japan*
[3]*Institute for Advanced Research, Nagoya University, Nagoya 464-8601, Japan*

We study polycrystalline samples of the hexagonal pnictides, CaAgP and CaAgAs, both of which are ideal candidates for line-node Dirac semimetals. The polycrystalline samples of CaAgP and CaAgAs obtained in this study are low-carrier metals, where hole carriers are dominant. By combining the hole carrier densities estimated from Hall coefficients and the electronic structures calculated by first principles calculations, both samples are found to have a ring-torus Fermi surface, derived from a ring-shaped Dirac line node. In the phosphide sample, the Fermi energy $E_F$ is located at around the end of the linear dispersion region of the electronic bands, while the $E_F$ in the arsenide sample exists in the middle of this region, suggesting that the arsenide is a more promising system for uncovering the physics of line-node Dirac semimetals.

In recent years, Dirac and Weyl semimetals, which are zero-gap semiconductors with linear-dispersion bands at the zero-gap points, have attracted broad interest as candidate systems for the realization of topologically nontrivial states in bulk materials.[1-8] In Dirac and Weyl semimetals, low-energy excitations of electrons at the zero-gap points are identical to those of relativistic massless Dirac and Weyl fermions, respectively. This is same as for graphene or the surface states of topological insulators. However, the Dirac and Weyl semimetals have three-dimensional linear dispersions, different from these two-dimensional systems, providing us with a chance of finding novel topological phenomena in condensed matter. Very recently, three-dimensional linear dispersions have been observed in photoemission spectra taken using single crystals of Na$_3$Bi and Cd$_3$As$_2$.[9-13] Transport measurements of these single crystals also yield anomalous behaviors, such as a gigantic linear magnetoresistance coming from the extremely high carrier mobility and a negative longitudinal magneto-resistance due to the chiral anomaly.[14,15] Moreover, surface Fermi arcs connecting the zero-gap points have been discovered in single crystals of TaAs and related compounds, suggesting the presence of the Weyl fermion in the bulk materials.[16-18]

In these Dirac and Weyl semimetals, the zero-gap points appear as discrete points in momentum space. In contrast, some systems are theoretically indicated to have a "line-node", where the linear dispersion bands cross on a line in momentum space.[19-30] These line-node Dirac and Weyl semimetals are expected to show anomalous electronic properties, such as a charge polarization and an orbital magnetization both proportional to the length of the line node, which do not appear in "point node" Dirac and Weyl semimetals.[31] The line node has thus far been experimentally observed by photoemission spectroscopy on single crystals of PbTaSe$_2$ and $M$SiS ($M$ = Zr or Hf).[32-36] However, the transport properties of these compounds are probably dominated by normal conduction electrons, because they have normal bands crossing the Fermi energy $E_F$ and their zero-gap points are shifted from $E_F$. Alternatively, a high temperature phase of Ca$_3$P$_2$ has been found to be an ideal line-node Dirac semimetal, where the only Dirac points exist at the $E_F$, by first principles calculations.[37] However, its physical properties have not been reported, mainly because of the difficulty in sample preparation.

In this letter, we report the physical properties of candidate line-node Dirac semimetals CaAg$X$ ($X$ = P or As). These compounds were firstly synthesized by Mewis in 1979 and reported to crystallize in the hexagonal ordered-Fe$_2$P type structure, as shown in Fig. 1(a).[38] This crystal structure has the space group of $P$−62$m$ (No. 189) with mirror planes instead of inversion centers. Although some ordered-Fe$_2$P type pnictides are of interest for their electronic properties such as superconductivity in ZrRuP ($T_c$ ~ 12 K) and the recently discovered ScIrP ($T_c$ = 3.4 K),[39,40] there is no report on the physical properties of CaAgP and CaAgAs.

Recently, we found that CaAgP and CaAgAs are ideal line-node Dirac semimetals by first principles calculations and theoretical analyses on the topological invariants.[41] First principles calculations without spin−orbit interactions indicate that the Dirac points form a ring at $k_z$ = 0, which is protected by the mirror symmetry located in the *xy* plane, as seen in the right panel of Fig. 1(a). Reflecting the topology of this "Dirac ring", dense surface states emerge inside the ring. When the spin−orbit interactions are switched on, CaAgP is still almost gapless (the size of the spin−orbit gap is less than 1 meV), while a spin−orbit gap of approximately 0.1 eV opens at the Dirac ring in CaAgAs due to the strong spin-orbit interactions of arsenic. This gapped state is found to be a topological insulator with the 1;000 $Z_2$ topological invariants.[41]

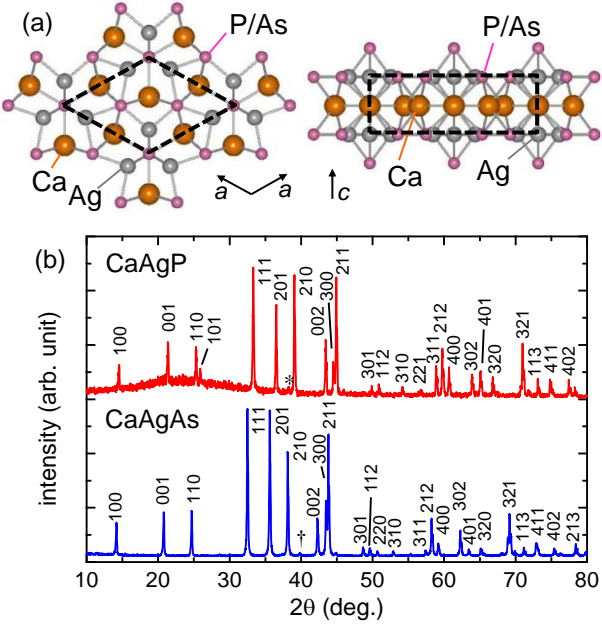

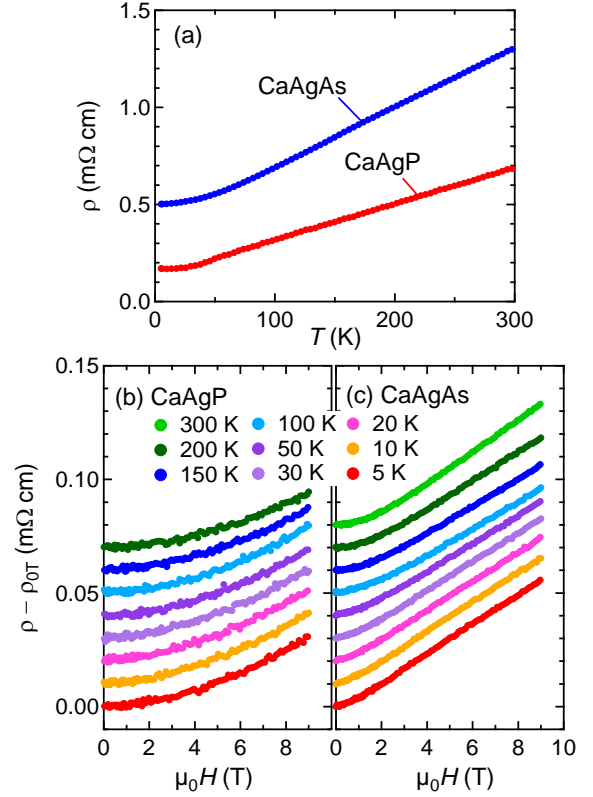

Fig. 1. (a) Crystal structure of CaAg$X$ ($X$ = P or As) viewed along (left) and perpendicular (right) to the $c$−axis. Large, medium, and small spheres represent Ca, Ag, and P/As atoms, respectively. The dashed lines represent the hexagonal unit cell. (b) Powder XRD patterns of CaAgP (upper) and CaAgAs (lower) polycrystalline samples taken at room temperature. The peaks indicated by an asterisk or a dagger are that of silver or an unknown impurity, respectively. Peak indices are given using hexagonal unit cells with lattice constants of $a$ = 7.0431(6) Å and $c$ = 4.1614(4) Å for CaAgP and $a$ = 7.2127(9) Å and $c$ = 4.2718(5) Å for CaAgAs.

An important feature of the Dirac semimetallic states in CaAgP and CaAgAs is that the Dirac ring lies solely at the $E_F$, different from almost all of the reported line-node Dirac semimetals. This feature results in a simple electronic structure, ideal for studying the physics of line-node Dirac semimetals. We succeeded in preparing polycrystalline samples of both compounds and measuring their physical properties. The hole carrier densities estimated from the Hall coefficients suggest that the $E_F$ of the CaAgP sample is located around the end of the linear dispersion region, while that of the CaAgAs sample is in the middle of the linear dispersion region, giving rise to a thin ring-torus Fermi surface. The arsenide sample is found to show a negligibly small Sommerfeld coefficient and linear magnetoresistance, reflecting the very small Fermi surface in the sample. These results indicate that CaAgAs is an ideal platform for uncovering the physics of line-node Dirac semimetals.

Fig. 2. (a) Temperature dependences of electrical resistivity of CaAgP and CaAgAs polycrystalline samples. Magnetic field dependences of electrical resistivity of (b) CaAgP and (c) CaAgAs polycrystalline samples measured at various temperatures. Magnetic fields are applied perpendicular to the electrical current. Resistivity data after subtraction of the zero-field data are indicated in (b) and (c). The data taken at 10–300 K are shifted for clarity.

We prepared polycrystalline samples of CaAgP and CaAgAs by a solid-state reaction method. An equimolar mixture of calcium chips, silver powder, and black phosphorus powder or arsenic chunks were put in an alumina crucible and sealed in an evacuated quartz tube. In the phosphide case, the tubes were kept at 673 K for 12 h and then at 1273 K for 3 h, followed by furnace cooling to room temperature. In the arsenide case, the tubes were kept at 773 K for 12 h and then 1323 K for 6 h and slowly cooled to 1083 K at a rate of 5 K h$^{-1}$, followed by furnace cooling to room temperature. The obtained as-grown samples were pulverized, pressed into pellets, and put in alumina crucibles sealed in quartz tubes. The pellets were sintered at 1073 K for 3 h and at 1173 K for 6 h for the phosphide and arsenide samples, respectively.

Sample characterization was performed by powder X-ray diffraction (XRD) analysis with Cu K$\alpha$ radiation at room temperature using a RINT-2100 diffractometer (Rigaku). As shown in Fig. 1(b), the diffraction peaks observed in the powder XRD patterns are sharp, indicative of the good crystallinity of the samples. All the peaks except for small peaks from impurities can be indexed on the basis of hexagonal structures with the lattice constants $a$ = 7.0431(6) and 7.2127(9) Å and $c$ = 4.1614(4) and 4.2718(5) Å for CaAgP and CaAgAs, respectively. The obtained $a$ and $c$ values are almost the same as those reported by Mewis.[38] Electrical resistivity, heat capacity, and Hall resistivity were measured using the Physical Properties Measurement System (Quantum Design). The as-grown samples were used for the electrical resistivity and Hall resistivity measurements of the phosphide samples, while the sintered samples were used for all other measurements. The phosphide samples were handled in an inert atmosphere.

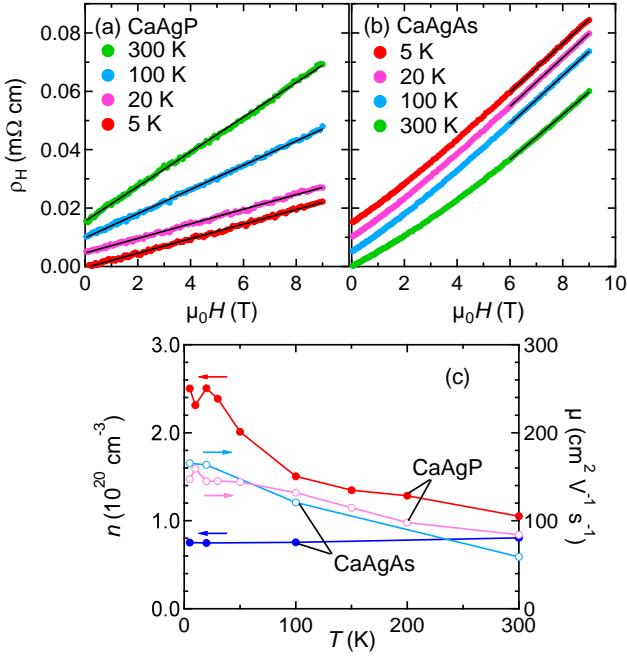

Fig. 3. Hall resistivity of (a) CaAgP and (b) CaAgAs polycrystalline samples measured in magnetic fields up to 9 T taken at 5, 20, 100, and 300 K and (c) hole carrier densities (filled circles) and mobilities (open circles) estimated from the Hall resistivity data. In (a) and (b), the data taken at 300, 100, and 20 K for CaAgP and at 5, 20, and 100 K for CaAgAs are shifted upward by 0.015, 0.01, and 0.005 mΩ cm, respectively.

Figure 2(a) shows the temperature dependences of the electrical resistivity, $\rho$, of polycrystalline samples of CaAgP and CaAgAs. Both samples show metallic behavior, where $\rho$ decreases with decreasing temperature. The $\rho$ of the phosphide sample is smaller than that of the arsenide sample. The residual resistivities, $\rho_0$, of the phosphide and arsenide samples are 0.17 and 0.50 mΩ cm and the residual resistivity ratios, RRR = $\rho_{300K}/\rho_0$, are 4.0 and 2.6, respectively. The larger $\rho_0$ in the arsenide sample may be related to the electron scattering at the grain boundaries and/or lower carrier density of the sample, as seen in Fig. 3(c), which is discussed later.

Figures 2(b) and (c) show the magnetic field, $H$, dependences of $\rho$ of the CaAgP and CaAgAs samples, respectively. The $\rho$ data after subtraction of the zero-field data are indicated. The phosphide sample exhibits an $H^2$ dependence, which is explained by the Lorentz force acting on the conduction electrons, in the whole temperature range. In contrast, the $H$ dependence of the arsenide sample is linear at a high $H$ region. This linear region expands on going to lower temperatures and is above 1 T at 5 K, as shown in Fig. 2(c). The magnetoresistance of the arsenide sample at $\mu_0 H$ = 9 T and $T$ = 5 K is 14% of the zero-field $\rho$. This is much smaller than those of the single crystals of Dirac semimetals $Cd_3As_2$ and $Na_3Bi$,[15,42] although the present samples are polycrystalline.

To estimate the carrier density and mobility contributing to the metallic behavior in both samples, we measured their Hall resistivity, $\rho_H$. As shown in Figs. 3(a) and (b), the $\rho_H$ values of both samples increase with increasing $H$ in the whole temperature range, resulting in positive Hall coefficients, $R_H$, in both samples. The positive $R_H$ suggests that hole carriers are dominant and $E_F$ values are located in the valence band below the Dirac point. The $H$ dependences of $\rho_H$ are different between the phosphide and arsenide samples, as seen in Figs. 3(a) and (b). The $\rho_H$ of the phosphide sample linearly increases with increasing $H$, while the arsenide sample shows concave-upward curves, which become more prominent at high temperatures. This concave-upward $H$ dependence may reflect the presence of the thermally excited electron carrier occupying the conduction band above the spin–orbit gap.

The $\rho_H$ data show linear behaviors in the whole magnetic field range and above 6 T for the phosphide and arsenide samples, respectively, coming from their hole carrier dominance. The linear fits of these data yield the hole carrier density, $n = 1/R_H e = \mu_0 H/\rho_H e$, as shown in Fig. 3(c). The value of $n$ = 7.50(3) × $10^{19}$ $cm^{-3}$ in the arsenide sample at 5 K is a considerably small value for a metallic sample, suggesting that the $E_F$ is located close to the Dirac point. As a result of the thermal excitation of conducting carriers, $n$ slightly increases with increasing temperature. Alternatively, $n$ of the phosphide sample is estimated to be 2.50(2) × $10^{20}$ $cm^{-3}$ at 5 K. This $n$ is also of the order of a low-carrier metal, but is more than three times larger than that of the arsenide sample. The $n$ of the phosphide sample decreases with increasing temperature, $n$ at 300 K is less than half of $n$ at 5 K, as shown in Fig. 3(c). The origin of this temperature dependence remains an open question, although it may reflect the fine structure of the electronic bands.

The hole carrier mobilities, $\mu = R_H/\rho$, of the CaAgP and CaAgAs polycrystalline samples are also shown in Fig. 3(c). The μ of both samples slightly decreases with increasing temperature and is approximately 100 cm$^2$ V$^{-1}$ s$^{-1}$ in the whole temperature range. This value is larger than that of a normal metal, such as Na and Cu with μ ~ 50 cm$^2$ V$^{-1}$ s$^{-1}$ at room temperature, but much smaller than those of Dirac semimetals, such as $Cd_3As_2$,[15] probably due to the electron scattering at grain boundaries in the polycrystalline sample.

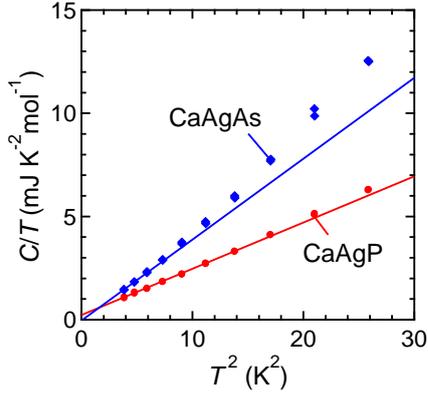

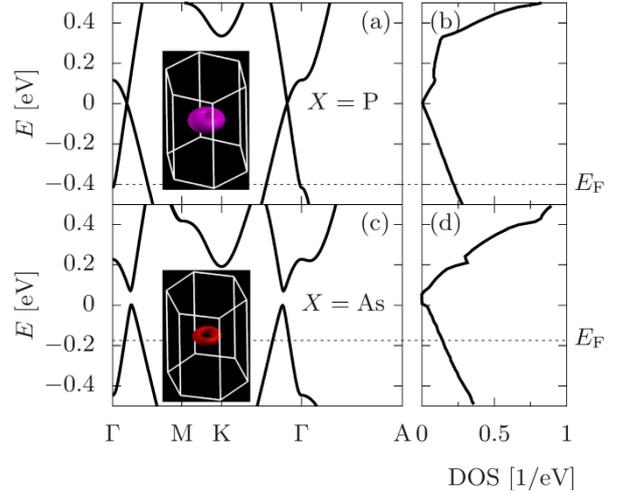

Fig. 4. Heat capacity divided by temperature of CaAgP and CaAgAs polycrystalline samples as a function of $T^2$. The solid lines show results of the linear fits between 2 and 4 K for CaAgP and 2 and 2.3 K for CaAgAs.

Heat capacity data also suggest that the CaAgP and CaAgAs samples are low-carrier metals. Figure 4 shows the heat capacity, $C$, divided by temperature of the CaAgP and CaAgAs polycrystalline samples as a function of $T^2$. Linear fits of the CaAgP and CaAgAs data below 4 and 2.3 K to the equation $C/T = AT^2 + \gamma$ yield $A$ = 0.224(2) and 0.39(11) mJ K$^{-4}$ mol$^{-1}$ and $\gamma$ = 0.22(17) and 0.04(5) mJ K$^{-2}$ mol$^{-1}$, respectively, where $A$ and $\gamma$ represent the coefficient of the $T^3$ term of the lattice heat capacity and the Sommerfeld coefficient, respectively. The γ values of both samples are close to zero, indicating that they have quite small electronic density of states at $E_F$.

We now discuss the electronic structures of the CaAg$X$ samples obtained in this study. First principles calculation results on CaAgP and CaAgAs are shown in Fig. 5. The Dirac ring in CaAgP and CaAgAs appears as a band crossing points on the Γ–K and Γ–M lines, where small spin–orbit gaps of less than 1 meV and of approximately 0.1 eV open in the phosphide and arsenide samples, respectively, as shown in Figs. 5(a) and (c). We also indicate the $E_F$ values calculated with experimentally obtained $n$ values at 5 K and the Femi surfaces with these $E_F$ values in Figs. 5(a) and (c). Both samples have a hole Fermi surface in the shape of a ring torus, derived from the Dirac ring. Such a ring-torus Fermi surface is quite rare and has not been reported thus far to our knowledge, although a spindle-torus one, where two circles overlap in a plane including the z axis, is realized as a spin-polarized Fermi surface in the giant-Rashba semiconductor BiTeI.[43,44] It would be interesting if superconductivity is discovered in CaAg$X$, because topological superconductivity, such as chiral $p$-wave superconductivity, can be clearly defined in metals with a ring-torus Fermi surface.

Fig. 5. Electronic states of (a, b) CaAgP and (c, d) CaAgAs with spin−orbit interactions. Electronic band structures (a, c) and electronic density of states (b, d) are shown. The tops of the valence bands are set to 0 eV. The Fermi energies and surfaces of the CaAgP and CaAgAs polycrystalline samples obtained in this study are also shown.

Fermi surfaces in the phosphide and arsenide samples are the same in topology, as discussed above, but their thicknesses are significantly different. The $n$ of the phosphide sample is estimated to be 2.5 × 10$^{20}$ cm$^{-3}$ at 5 K, which is approximately three times larger than that of the arsenide sample. This $n$ yields the $E_F$ location below 0.4 eV from the top of the valence band, as shown in Fig. 5(a), close to the end of the linear dispersion region. In this case, the Sommerfeld coefficient is calculated to be $\gamma_{band}$ = 0.2 mJ K$^{-2}$ mol$^{-1}$, which is almost identical to the experimental value of $\gamma$ = 0.22(17) mJ K$^{-2}$ mol$^{-1}$. In contrast, as shown in Fig. 5(c), the $E_F$ of the arsenide sample estimated from $n$ = 7.5 × 10$^{19}$ cm$^{-3}$ is located 0.2 eV below the top of the valence band, which is in the middle of the linear dispersion region. This $n$ gives rise to a $\gamma_{band}$ of 0.1 mJ K$^{-2}$ mol$^{-1}$, smaller than that of the phosphide sample, which coincides with the smaller $\gamma$ = 0.04(5) mJ K$^{-2}$ mol$^{-1}$ in the arsenide sample than $\gamma$ = 0.22(17) mJ K$^{-2}$ mol$^{-1}$ of the phosphide sample. Therefore, the Fermi surface of the arsenide sample is much thinner than that of the phosphide sample, as seen in Figs. 5(a) and (c).

This difference in thickness of the ring-torus Fermi surfaces indicates that the arsenide sample has the electronic state in the vicinity of the line-node Dirac semimetal compared to the phosphide sample, which is expected to have significant influence on the physical properties. For example, it may give rise to different $H$ dependences of ρ between the arsenide and phosphide samples seen in Figs. 2(b) and (c), respectively. However, it is important to note that the arsenide sample obtained in this study is not perfect for uncovering the nature of line-node Dirac semimetals, because the sample is polycrystalline and hole doped probably due to lattice defects. Synthesis of the single crystals with the reduced $n$ of the order of $10^{18}$ cm$^{-3}$ comparable to Cd$_3$As$_2$ will allow us to find unique physical properties originating from the Dirac line node.

In summary, we prepared polycrystalline samples of CaAgP and CaAgAs, which are promising candidates for ideal line-node Dirac semimetals. CaAgP and CaAgAs samples are found to be $p$-type metals with hole carrier densities of $n = 2.5 \times 10^{20}$ and $7.5 \times 10^{19}$ cm$^{-3}$, respectively, by resistivity and Hall resistivity measurements. On the basis of these $n$ values and first principles calculations, the $E_F$ in the phosphide sample is determined to be around the end of the linear dispersion region, while the $E_F$ of the arsenide sample is in the middle of this region. The CaAgP and CaAgAs samples have a ring-torus Fermi surface, reflecting the Dirac ring, and the arsenide has a much thinner torus than that of the phosphide, suggesting that CaAgAs is an excellent platform for studying the physics of line-node Dirac semimetals.

**Acknowledgments**

This work was partly carried out at the Materials Design and Characterization Laboratory under the Visiting Research Program of the Institute for Solid State Physics, University of Tokyo and partly supported by the Collaborative Research Project of Materials and Structures Laboratory, Tokyo Institute of Technology, JSPS KAKENHI Grant Number 16K13664, and Iketani Science and Technology Foundation.

______________________________________________________

*yokamoto@nuap.nagoya-u.ac.jp